\documentclass[12pt,preprint]{aastex}
\usepackage{lscape}

\begin{document}
\newcommand{\bdv}[1]{\mbox{\boldmath$#1$}}
\def\kms{{\rm km\,s^{-1}}}
\def\esc{{\rm esc}}
\def\circ{{\rm circ}}
\def\cut{{\rm cut}}
\def\tot{{\rm tot}}
\def\thresh{{\rm thresh}}
\def\bv{{\bf v}}
\def\rel{{\rm  rel}}
\def\bmu{{\bdv{\mu}}}
\def\bpi{{\bdv{\pi}}}
\def\au{{\rm AU}}
\def\e{{\rm E}}

\title{Microlens Parallax Measurements with a Warm {\it Spitzer}}
\author{Andrew Gould}
\affil{
Department of Astronomy, The Ohio State University, 
140 West 18th Avenue, Columbus, OH 43210\\
gould@astronomy.ohio-state.edu
}

\begin{abstract}

Because {\it Spitzer} is an Earth-trailing orbit, losing about 0.1 AU/yr,
it is excellently located to perform microlens parallax observations
toward the Magellanic Clouds (LMC/SMC) and the Galactic bulge.  These yield
the so-called ``projected velocity'' of the lens, which can distinguish
statistically among different populations.  A few such measurements
toward the LMC/SMC would reveal the nature of the lenses being detected
in this direction (dark halo objects, or ordinary LMC/SMC stars).
Cool {\it Spitzer} has already made one such measurement of a (rare)
bright red-clump source, but warm (presumably less oversubscribed) 
{\it Spitzer} could devote the extra time required to obtain microlens
parallaxes for the more common, but fainter, turnoff sources.
Warm {\it Spitzer} could observe bulge microlenses for 38 days per year,
which would permit up to 24 microlens parallaxes per year.  This
would yield interesting information on the disk mass function, particularly
old brown dwarfs, which at present are inaccessible by other techniques.
Target-of-Opportunity (TOO) observations should be divided into RTOO/DTOO,
i.e., ``regular'' and ``disruptive'' TOOs, as pioneered by 
the {\it Space Interferometry Mission (SIM)}.  LMC/SMC  parallax measurements
would be DTOO, but bulge measurements would be RTOO, i.e., they could be
scheduled in advance, without knowing exactly which star was to be observed.
\end{abstract}

\section{{Introduction}
\label{sec:intro}}

Microlens parallaxes measure a vector quantity, the ``projected velocity''
$\tilde\bv$, which is the projection of the lens-source relative
velocity on the plane of the observer.  Another way of writing this
quantity is 
\begin{equation}
\tilde \bv = \au{\bmu_\rel\over\pi_\rel}.
\label{eqn:vtilde}
\end{equation}
where $\pi_\rel$  and $\bmu_\rel$ are the lens-source relative parallax
and proper motion.  It is a useful quantity to measure because
it depends only on the kinematic properties of the lens and source,
and is independent of the mass. Once it is measured, one can also
determine the ``projected Einstein radius",
\begin{equation}
\tilde r_\e = \tilde v t_\e = \sqrt{\kappa M\over\pi_\rel},
\label{eqn:retilde}
\end{equation}
where $t_\e$ is the ``Einstein timescale" (which is almost always
well-measured) and $\kappa\equiv 4 G /c^2\au = 8.1\,{\rm mas}/M_\odot$.

Hence, an ensemble microlens parallax measurements can distinguish
statistically between different kinematic populations.

\section{{Microlens Parallax Science with a Warm {\it Spitzer}}
\label{sec:science}}

Microlensing parallax measurements are feasible with {\it Spitzer}
toward two classes of targets, the Large and Small Magellanic Clouds
(LMC and SMC) and the Galactic bulge.  The science cases and
technical challenges are substantially different for the two classes.

\subsection{{Science toward the LMC/SMC}
\label{sec:lmcscience}}

The nature of the microlensing events detected toward the Magellanic
Clouds by the MACHO \citep{alcock00}and EROS \citep{tisserand07}
collaborations is unknown.  They
are definitely too infrequent to make up most of the dark matter,
but could make up of order 20\%.  The basic problem is that, while
a few of the lenses have special properties that allow their
distances to be determined, for most lenses it is not known
whether they are in the Milky Way (MW) halo, the MW disk, or
in the Clouds themselves.  While it would be tempting to use
the lenses with known locations to address this question, actually
the very properties that allow their distances to be measured
predispose them to be in the Clouds or the MW, so they do not
constitute a fair sample.  It would be better to find a way
to choose ``typical" events, not selected by any characteristic,
and determine their position.

{\it Spitzer} has already been used to measure the microlens
parallax of one lens.  \citet{dong07} found that its projected
velocity was $\tilde v \sim 230\,\kms$, which is typical of
halo lenses, but much bigger than expected for MW disk lenses
and much smaller than expected for SMC lenses.  However, they
also found that (with no priors on the location or properties of the
lens) that it could be in the SMC with 5\% probability.  Thus,
to draw rigorous scientific conclusions, it will be necessary to
obtain parallaxes for at least 3, and more comfortably 5, such
events.

\subsection{{Science toward the Galactic bulge}
\label{sec:bulgescience}}

\citet{han95} conducted a systematic analysis of what could be learned
about lenses observed toward  the bulge by obtaining parallaxes
for an ensemble of them.  They found that such parallaxes greatly
enhance microlensing as a probe of the stellar mass function.
For example, comparing the precision of mass estimates (for lenses
assumed to be in the disk in both cases), they found that without
parallaxes the masses could be individually ``measured" with a $1\,\sigma$
precision of 0.6 dex (so, essentially no measurement) whereas with
parallaxes, the uncertainty was reduced to 0.2 dex.  

Of particular interest would be to measure the frequency of 
disk brown dwarfs (BDs).  Of course, BDs can be directly detected,
but only while they are young, so that inferences about their
global  frequency depend sensitively on both their history of formation
and their assumed cooling properties.  It would be nice to constrain
their frequency based solely on their mass.

The main characteristic of disk BDs from a parallax standpoint is
that they have small $\tilde r_\e$ (see eq.~[\ref{eqn:retilde}]).  
For example, a BD with $M=0.03\,M_\odot$ and distance $D_l=3\,$kpc
would have $\tilde r_\e = 1.1\,\au$.  In order for an M dwarf
with mass $M=0.08\,M_\odot$ to produce the same $\tilde r_\e$ would
require a distance $D_l = 1.5\,$kpc.  This is much less likely,
but of course not impossible.  So an ensemble of measurements would be
needed to estimate to estimate the BD frequency.
In fact, as I will discuss
in the \S~\ref{sec:bulgetech}, parallax measurements are actually
easier for lenses with smaller  $\tilde r_\e$ (up to a point).
So {\it Spitzer} is well-placed to probe this key regime.

\section{{Technical Challenges}
\label{sec:technical}}

The technical challenges in the two directions differ significantly,
although they are related.

\subsection{{Technical Challenges toward the Clouds}
\label{sec:lmctech}}

As mentioned, there is already one {\it Spitzer} microlens parallax toward
the Clouds, which is a proof of concept.  However, the source was
an $I=18$ clump giant (i.e., quite cool, so large $I-L$), 
which is what permitted
excellent photometry with just 2 hours of integration
time in each of the four measurements required.  Such microlensing sources 
toward the Clouds are extremely rare (this was the only one detected so 
far during the entire {\it Spitzer} mission).  To obtain the 3--5
events that are needed for the science goal, requires acceptance of
more typical turnoff stars, which are both fainter and bluer, and hence
demand integration times that are 10--20 times longer.  Such
integrations were prohibitive for cold {\it Spitzer} but might
be possible for warm {\it Spitzer} when the observatory is under
less severe demand.  If we imagine an average of 60 hours per event
for 4 events, this would take about 250 hours over 4 years.
Note that the individual exposures cannot be made any longer than
the 27 second exposures for the clump giant because of artifacts
from very bright stars.

\subsection{{Technical Challenges toward the Bulge}
\label{sec:bulgetech}}

Here there are many challenges.  First, because the bulge
is near the ecliptic, it can only be observed for two 38-day periods
per year (in late spring and late autumn).  Moreover, the autumn
window is almost useless for microlensing because no events can be
discovered then from the ground.  Second, in contrast to the Clouds,
a large number of events must be monitored to obtain 
scientifically interesting results, close to 50 and preferably 100.
Third, there are technical challenges related to reconciling
ground-based and {\it Spitzer} photometry.  This is a classic
problem first discussed by \citet{gould95}.  Basically, unlike
trigonometric parallax, microlens parallax is actually a vector,
$\bpi_\e$, which is related to observations by
\begin{equation}
\bpi_\e = {\au\over d_{\rm proj}}
\biggl({\Delta t_0\over t_\e},\Delta u_0\biggr)
\label{eqn:bpie}
\end{equation}
where $\Delta t_0$ is the measured difference in times of peak of the
event as seen from {\it Spitzer} and the ground, $\Delta u_0$ is
the measured difference in the impact parameter,  and $d_{\rm proj}$
is the magnitude of the Earth-{\it Spitzer} separation vector
projected onto the plane of the sky.  The problem is that while
$\Delta t_0$ can be robustly measured (because the time of the
peak can be read directly from the lightcurve), $u_0$ from each
observatory is much harder to measure because it is a fit parameter
that is partially degenerate with the amount of blended light.
\citet{gould95} showed this degeneracy could be strongly constrained
by arranging for the space and ground observatories to have
identical cameras (and so identical blending) but that is obviously
impossible for {\it Spitzer} observations.  \citet{dong07} evaded
this problem by making {\it Spitzer} responsible only for
measuring the first component of $\bpi_\e$: the events toward the
Clouds are long enough that the acceleration of the Earth toward
the Sun (roughly perpendicular to {\it Spitzer}) allowed ground-based
measurement of the other component.  However, most  events  toward
the Galactic bulge are too short to make this trick work.

\begin{figure}
\plotone{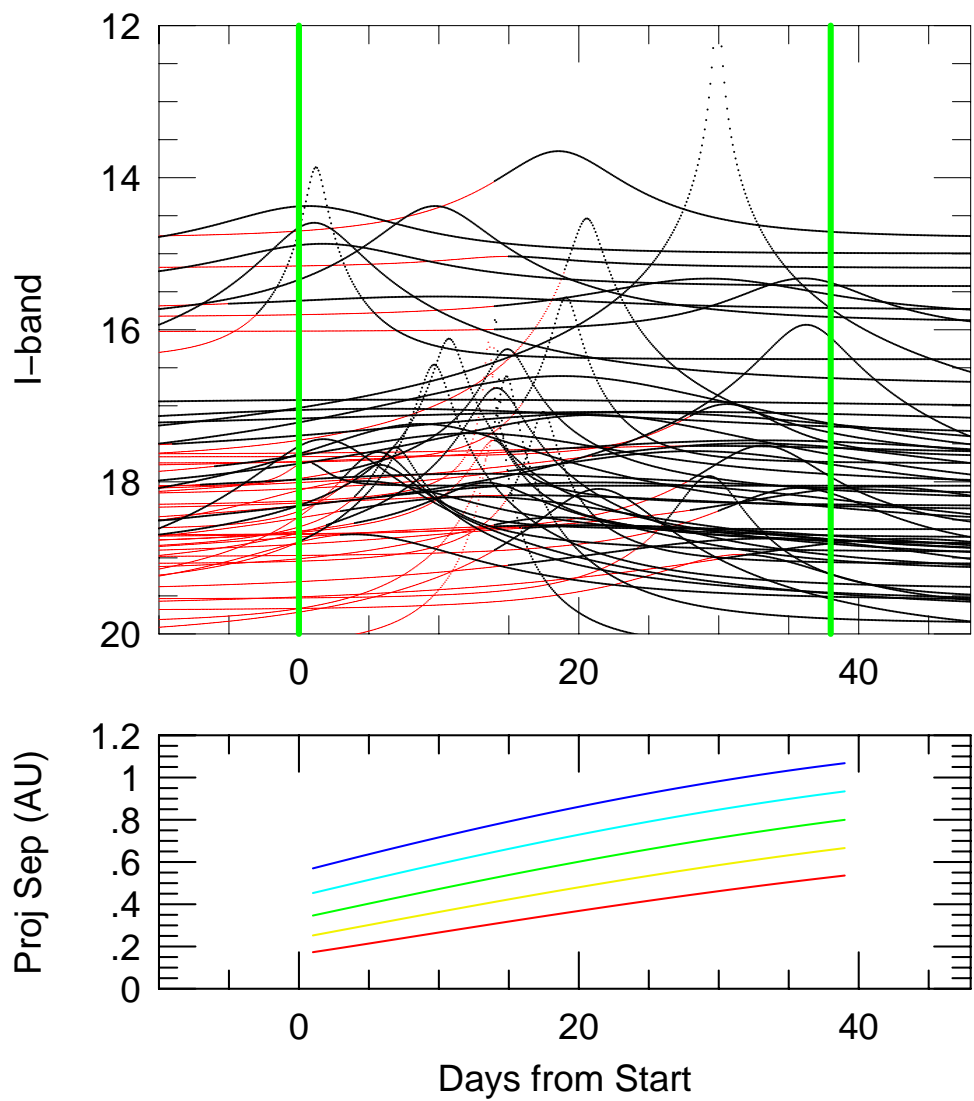}
\caption{\label{fig:ogle_poss}
Upper Panel: Light curve trajectories of all 66 events discovered 
before peak by
the OGLE collaboration \citep{ews} during an arbitrarily chosen
38-day period beginning 1 April 2007.  Thirty-eight day interval
is delineated by green vertical lines.  Light curves are in red
prior to discovery of event and in black afterward.  Lower Panel:
Earth-{\it Spitzer} separation (projected onto the plane of the
sky toward the Galactic bulge) as a function of time for each
of the five 38-day intervals that warm {\it Spitzer} will be
able to observe the bulge.  Red to blue corresponds to 2009 to 2013.
%Same as Figure \ref{fig:ogle_all}, except restricted to the 66
%events that were alerted by OGLE before peak.
}\end{figure}

%In order to understand whether the first two challenges can be
%overcome, I plot in Figure \ref{fig:ogle_all}, all 90
%OGLE events that peaked during a random 38-day period in Spring 2007.
%The events are plotted in red before they were alerted and
%in black afterward.  Events must be alerted  before peak or
%it is impossible to measure $t_0$ from {\it Spitzer}.  The
%diagram is quite a mess, but much of this mess is caused by
%24 events that were alerted after peak because they are so faint.
%The remaining 66 events are shown in Figure \ref{fig:ogle_poss}.  Based
In order to understand whether the first two challenges can be
overcome, I plotted (not shown), all 90
OGLE events that peaked during a random 38-day period in Spring 2007, with
events plotted in red before they were alerted and
in black afterward.  Events must be alerted  before peak or
it is impossible to measure $t_0$ from {\it Spitzer}.  The
diagram proved to be quite a mess, but much of this mess is caused by
24 events that were alerted after peak because they are so faint.
The remaining 66 events are shown in Figure \ref{fig:ogle_poss}.  Based
on experience with {\it Spitzer} observations of the SMC event,
it should be possible to obtain better than 1\% photometry in
one  hour exposures of events that reach $I<17$.  (The bulge
sources are intrinsically bluer than the SMC clump giant,
but the fields are heavily reddened, so the {\it observed}
$I-L$ colors are similar.)  There are 24 such events.
Hence, over 5 years, a sample of 120 events could yield measured
parallaxes.

Before addressing the challenge of aligning the photometry of the
two observatories, one should take note of the bottom panel in
Figure \ref{fig:ogle_poss}.  
It shows the Earth-{\it Spitzer} separation projected
onto the bulge, during each of the 5 38-day intervals of the
Warm {\it Spitzer} Mission, red for the first year and blue for
the last.  The average value is about 0.6 AU, about 2.5 times the leverage
available for the SMC event \citep{dong07}.  This is important because, until
the projected separation gets to be the same order as $\tilde r_\e$,
the signal-to-noise ratio of the measurement scales directly as
the projected separation.

A possible approach to aligning the ground and {\it Spitzer}
photometry would be to obtain $L$ band images during the event
from the ground with a high-resolution, large aperture telescope.
Such measurements would yield an accurate $I-L_{\rm ground}$ color, 
independent of blending.  While $L_{\rm ground}$ is certainly not
identical to {\it Spitzer} $3.6\,\mu$m, it should be possible
to use this color, together with the colors of other stars
in the image, to align the two photometry systems based on
stars of similar color.  Only empirical testing will determine
whether this is a practical method.  One potential problem with
this approach is that there may not be enough bright bluish stars
to perform the alignment for the (typically bluish turnoff) stars
that dominate the event rate.  In this case, one might be forced
to stick with the brighter stars, which comprise 7 stars in the
interval shown in Figure \ref{fig:ogle_poss}.

Even if it proves difficult to measure $\Delta u_0$ very accurately,
it is important to point out that for one of the most interesting
applications, the BD mass function, $\Delta u_0$ does not have
to be measured with great precision.  With typical $\tilde r_\e\la 2\,\au$
and $d_{\rm proj}\sim 0.6\,\au$, measurements of $\Delta u_0$
accurate to 0.1  would be quite adequate.

\subsection{{RToO/DToO}
\label{sec:r2d2}}

Finally, there is the question of the practical organization
of bulge observations.  Target-of-Opportunity (ToO) observations
are notoriously disruptive, and the Warm {\it Spitzer} Mission
is likely to have a lower level of staff support to deal with
these disruptions than the cold mission does.

However, given the very large number of observations required
during each 38-day observing ``season", it should be possible
apply the concept [originally developed for the {\it Space Interferometry
Mission (SIM)}] of a ``Regular ToO" or RToO (as opposed to a
``Disruptive ToO" or DToO).  For an RToO, one plans in advance
to make observations of the bulge for a designated set of times,
but without specifying exactly what the targets are.  Then
a day or so before the observations, instructions are uploaded to
the spacecraft specifying the targets.

If it really proves possible to measure 24 targets, and each one
requires 7 1-hr measurements (plus 1 baseline measurement that can
be done at leisure during the autumn 38-day interval), then
this amounts to 168 hours of observing over 38 days, or about
18\% of observing time during this interval.  All of these
observations could be scheduled in advance, without knowing
what the targets were.

%\begin{equation}
%\label{eqn:}
%\end{equation}

%\begin{equation}
%\label{eqn:}
%\end{equation}

{}
\eject

%\begin{figure}
%\plotone{events_all.ps}
%\caption{\label{fig:ogle_all}
%}\end{figure}

\end{document}